\documentclass[prb,twocolumn,showpacs,superscriptaddress,floatfix,longbibliography]{revtex4-1}
\usepackage{graphicx,amsfonts,amssymb,amsmath,xspace,dsfont}
\usepackage[colorlinks=true,citecolor=green,linkcolor=red,urlcolor=blue]{hyperref}
\usepackage{bbm}

\usepackage{color}
\definecolor{g}{rgb}{.1,0.4,.1} % {.0,0.7,.5}
\definecolor{b}{rgb}{0,0.2,1}
\definecolor{rouge}{rgb}{0.82,0.,0.}
\definecolor{vert}{rgb}{0.,0.82,0.}
\definecolor{orange}{rgb}{1,0.5,0.}
\definecolor{bleu}{rgb}{0.,0.,0.82}
\definecolor{m}{rgb}{0.82,0.,0.82}
\definecolor{vert2}{rgb}{0.,0.5,0.}
\definecolor{rougeclair}{rgb}{1.0,0.7,0.7}

\newcommand{\bra}[1]{\langle#1|}
\newcommand{\ket}[1]{|#1\rangle}
\newcommand{\braket}[1]{\langle#1|#1\rangle}
\newcommand{\mN}{\mathcal{N}}
\newcommand{\Np}{N_\mathrm{p}}
\newcommand{\mC}{\mathcal{C}}
\newcommand{\mD}{\mathcal{D}}

\newcommand{\id}{\mathbbm{1}}
\newcommand{\tr}{\mathrm{Tr}}
\newcommand{\Pp}{B_p}
\newcommand{\Pv}{Q_v}
\newcommand{\Pe}{L_l}

\begin{document}

\title{Mean-field ansatz for topological phases with string tension}

\author{S\'{e}bastien Dusuel}
\email{sdusuel@gmail.com}
\affiliation{Lyc\'ee Saint-Louis, 44 Boulevard Saint-Michel, 75006 Paris, France}
\author{Julien Vidal}
\email{vidal@lptmc.jussieu.fr}
\affiliation{Laboratoire de Physique Th\'eorique de la Mati\`ere Condens\'ee,
CNRS UMR 7600, Universit\'e Pierre et Marie Curie, 4 Place Jussieu, 75252
Paris Cedex 05, France}

%%%%%%%%%%%%%%%%%%%%%%%%%%%%%%%%%%%%%%%%%%%%%%%%%%%%%%%%%%%%%%%%%%%%%%%%%%%%%%%

\begin{abstract}
We propose a simple mean-field ansatz to study phase
transitions from a topological phase to a trivial phase. We probe the efficiency
of this approach by considering the string-net model in the presence of a string
tension for any anyon theory. Such a perturbation is known to be responsible for
a deconfinement-confinement phase transition which is well described by the present 
variational setup. We argue that  mean-field results become exact in the limit of large total quantum dimension.
\end{abstract}

\pacs{05.30.Pr, 05.30.Rt, 71.10.Pm, 75.10.Jm}

\maketitle

%%%%%%%%%%%%%%%%%%%%%%%%%%%%%%%%%%%%%%%%%%%%%%%%%%%%%%%%%%%%%%%%%%%%%%%%%%%%%%%
\section{Introduction}
%%%%%%%%%%%%%%%%%%%%%%%%%%%%%%%%%%%%%%%%%%%%%%%%%%%%%%%%%%%%%%%%%%%%%%%%%%%%%%%
The blend of quantum computation and of topological phases of matter
\cite{Wen13} have led to the idea of topological quantum computation
\cite{Kitaev03,Preskill_HP,Wang_book}. In this field, the essential ingredient
is the construction of physical systems sustaining exotic excitations known
as non-Abelian anyons (see Ref.~\cite{Nayak08} for a review). Being genuinely
nonlocal, these anyons allow for efficient storage and manipulation of quantum
information. Indeed, topologically ordered systems \cite{Wen13} are stable under
local perturbations \cite{Bravyi10} and hence protected against undesirable
effects such as decoherence. However, strong enough perturbations, may drive the
system to a nontopological phase. 
In recent years, many works have been devoted to the study of this robustness in microscopic models. Such an issue is difficult to address since one has to deal with two-dimensional interacting quantum systems and the complex nature of the anyonic quasiparticles prevents one from using standard methods. 

The goal of the present work is to propose a simple approach that may be
considered as a mean-field theory for topological phases. To this end, we
introduce a variational ansatz which can describe topological as well as non
topological phases. By construction, it also matches the exact ground state in
some limiting cases. Thus, it aims at qualitatively describing phase diagrams
while being quantitatively acceptable. Most models hosting topological quantum order  are built as a sum of local commuting projectors (toric code \cite{Kitaev03}, string nets  \cite{Levin05},...). In lattice gauge theories, one often interprets these projectors as operators measuring effective fluxes and charges.  The topologically ordered ground state (vacuum) is then defined as the  flux-free and charge-free state. Elementary excitations are obtained by locally violating this constraint. In two dimensions, excitations are pointlike anyons related by strings and
their energy does not depend on their relative position so that topological
phases are also called deconfined phases.  A natural way to destroy topological
order consists in adding a string tension that will drive the system to a confined phase. 
The prototypical Hamiltonian of such a system can be written
%
%
%%%%%%%%%%%%%%%%
\begin{equation}
H= -J_\mathrm{v}\sum_v \Pv- J_\mathrm{p}\sum_p \Pp -J_\mathrm{l}\sum_l\Pe,
\label{eq:ham}
\end{equation}
%%%%%%%%%%%%%%%%
%
%
where $\Pv$ ($\Pp$) are projectors measuring  charges (fluxes) on vertices
(plaquettes) of a two-dimensional graph and where $L_l$ is an operator acting on links which induces a string tension. 
In the following, we consider a two-dimensional plane with open boundary
conditions so that the ground state is unique in the thermodynamical limit.
Assuming non-negative couplings, the ground state of $H$ is readily written in
two limiting cases. On one hand, in the trivial phase
$J_\mathrm{v}=J_\mathrm{p}=0$, the ground state is a (polarized) product state
denoted by $\ket{0}$, where all links are in the same state. On the other hand,
for  $J_\mathrm{l}=0$, the ground state is proportional to $\prod_v \Pv \prod_p
\Pp \ket{0}$. The main idea of our construction is to find a simple variational
state that bridges the gap between these two extreme cases.

In this paper, we focus on the string-net model in the honeycomb lattice since
it allows one to study a wide variety of topological phases \cite{Levin05,Lin14}.
Interested readers that are not familiar with this model can find a detailed
study of this variational approach in the simpler case of the toric code model
in Appendix \ref{app_TCF}.

The string-net Hamiltonian \cite{Levin05} is a special case of
Eq.~(\ref{eq:ham}) where the operator $\Pp$ favors the zero-flux configuration
in plaquette $p$. Here, we only consider states without charge excitation so
that the Hilbert space is spanned by all link configurations satisfying the
so-called branching rules (stemming from the fusion rules of the considered
anyon theory). We thus drop the $-J_\mathrm{v}\sum_v \Pv$ term in the
Hamiltonian. For simplicity, we also restrict our discussion to the string
tension term introduced in Ref.~\cite{Gils09_1} which involves $\Pe$ operators
enforcing a zero flux in link $l$ of the lattice. Operators $\Pp$ and $\Pe$
commute except if link $l$ belongs to plaquette $p$.

%
%
%%%%%%%%%%%%%%%%%%%%%%%
\begin{figure}[t]
\includegraphics[width=0.7\columnwidth]{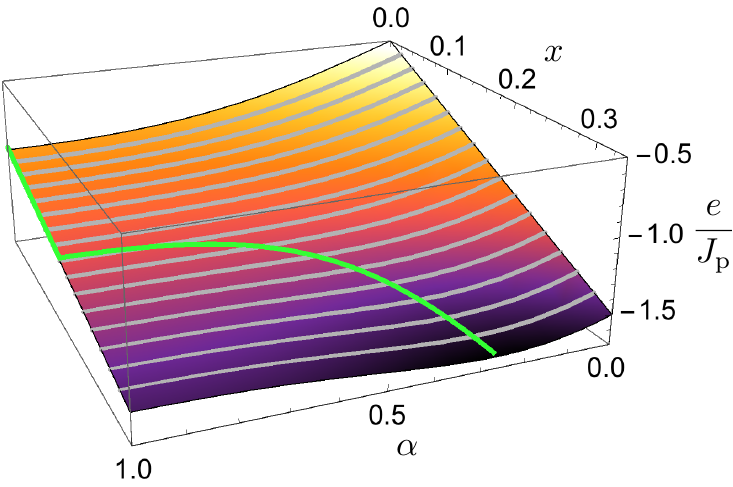}
\caption{(Color online) Energy landscape as a function of $\alpha$ and $x=J_\mathrm{l}/J_\mathrm{p}$ for $D^2=2$. The green line shows the position of the absolute minimum $\alpha(x)$. In this case, the transition is found to be continuous (second order). }
\label{fig:e1}
\end{figure}
%%%%%%%%%%%%%%%%%%%%%%%
%
%

%%%%%%%%%%%%%%%%%%%%%%%%%%%%%%%%%%%%%%%%%%%%%%%%%%%%%%%%%%%%%%%%%%%%%%%%%%%%%
\section{Ansatz state and its basic properties}
%%%%%%%%%%%%%%%%%%%%%%%%%%%%%%%%%%%%%%%%%%%%%%%%%%%%%%%%%%%%%%%%%%%%%%%%%%%%%%%
To describe the phase transition separating the topological phase from the trivial phase, we introduce the following single-parameter variational state:
%
%%%%%%%%%%%%%%%%
\begin{equation}
\ket{\alpha}=\mN\prod_p(\id+\alpha Z_p)\ket{0},
\label{eq:state}
\end{equation}
%%%%%%%%%%%%%%%%
%
where $0\leqslant\alpha\leqslant 1$, and $Z_p=2\Pp-\id$ is such that \mbox{$Z_p^2=\id$}.  
The normalization constant $\mN$ depends on the total quantum dimension $D$ of
the theory considered, on~$\alpha$, and on the system size (see Appendix \ref{app_SN}).
Once again, the physical insight underlying this ansatz is that $\ket{\alpha=0}=\ket{0}$ is the exact ground state for $J_\mathrm{p}=0$, while $\ket{\alpha=1}\propto\prod_p \Pp \ket{0}$ is the exact ground state for $J_\mathrm{l}=0$. Thus, one can expect that it captures the physics, at least qualitatively, for nonvanishing couplings. 

Interestingly, the structure of  $\ket{\alpha}$ implies that for any set $\mathcal{P}_n$ of $n$ plaquettes, one has 
%
%%%%%%%%%%%%%%%%
\begin{equation}
\Big\langle \prod_{p \in \mathcal{P}_n}  \Pp \Big\rangle_\alpha =\prod_{p \in \mathcal{P}_n} \langle \Pp \rangle_\alpha=\left[\frac{(1+\alpha)^2}{D^2(1-\alpha)^2+4\alpha}\right]^n,
\label{eq:factorization}
\end{equation}
%%%%%%%%%%%%%%%%
%
where $\langle\mathcal{O}\rangle_\alpha=\bra{\alpha}\mathcal{O}\ket{\alpha}$ (see Appendix \ref{app_SN} for details).
This factorization property reveals the mean-field character of $\ket{\alpha}$.
In addition, for Abelian theories, $\ket{\alpha}$ can be rewritten as a simple
product state in the dual plaquette (flux) basis.
For illustration, let us consider the simplest Abelian theory, i.e., $\mathbb{Z}_2$ ($D^2=2$). As shown in Ref.~\cite{Burnell11_2}, for this theory, the string-net model with a string tension can be mapped onto the transverse-field Ising model on the triangular lattice by setting \mbox{$X_p X_{p'}=2\Pe-\id$}, where $p$ and $p'$ are plaquettes sharing link $l$. 
In this dual representation, degrees of freedom are defined on plaquettes (instead of links) and operators $X_p$ and $Z_p$ are the usual Pauli matrices. 
One can then compute the following expectation values in the link basis (see Appendix \ref{app_SN})
%
%%%%%%%%%%%%%%%%
\begin{equation}
\langle 2\Pp-\id\rangle_\alpha=\frac{2\alpha}{1+\alpha^2} , \quad
\langle 2\Pe-\id\rangle_\alpha=\left(\frac{1-\alpha^2}{1+\alpha^2}\right)^2,
\label{eq:means}
\end{equation}
%%%%%%%%%%%%%%%%
%
and in the plaquette basis
%
%%%%%%%%%%%%%%%%
\begin{equation}
\langle Z_p\rangle_\theta=\cos\theta , \quad \langle X_pX_{p'}\rangle_\theta=\sin^2\theta=\langle X_p\rangle_\theta\langle X_{p'}\rangle_\theta.
\label{eq:means2}
\end{equation}
%%%%%%%%%%%%%%%%
%
Here, we set \mbox{$|\theta \rangle=\otimes_p \big[\cos (\theta/2)
\ket{\!\!\uparrow}_p+\sin (\theta/2) \ket{\!\!\downarrow}_p\big]$} where
$\ket{\!\!\uparrow}_p$ and  $\ket{\!\!\downarrow}_p$ are the eigenstates of
$Z_p$ with eigenvalues $+1$ and $-1$. Clearly, expressions (\ref{eq:means}) and
(\ref{eq:means2}) coincide provided $\alpha=\tan (\theta/2)$.

The $\mathbb{Z}_N$ case can be treated similarly by mapping the model onto the
transverse-field $N$-state Potts model \cite{Burnell11_2} (other models with
$\mathbb{Z}_2$ and $\mathbb{Z}_3$ topological order have also been treated in the same vein
\cite{Schulz12,Karimipour13, Mohseninia15}).
Although no such mapping is known for non-Abelian theories (because of the
existence of multiple fusion channels), the state $\ket{\alpha}$ can still be
considered as a mean-field ansatz because of the factorization
property~(\ref{eq:factorization}). 
In other words, the present approach generalizes the canonical mean-field treatment (performed in the dual basis) implemented for Abelian anyons, to non-Abelian theories. 

%
%
%%%%%%%%%%%%%%%%%%%%%%%
\begin{figure}[t]
\includegraphics[width=0.7\columnwidth]{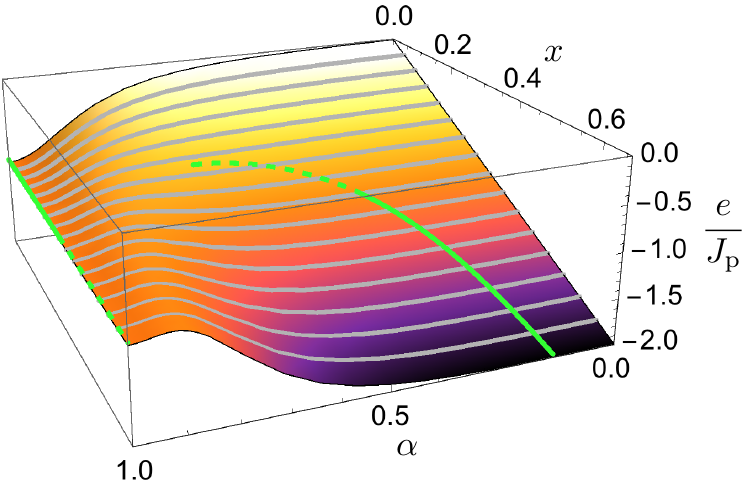}
\caption{(Color online) Energy landscape as a function of $\alpha$ and $x=J_\mathrm{l}/J_\mathrm{p}$ \mbox{for $D^2=100$}. The green (dotted) line shows the position of the absolute (local) minimum $\alpha(x)$. For \mbox{$D^2>2$}, the transition is found to be discontinuous (first order).}
\label{fig:e2}
\end{figure}
%%%%%%%%%%%%%%%%%%%%%%%
%
%

%%%%%%%%%%%%%%%%%%%%%%%%%%%%%%%%%%%%%%%%%%%%%%%%%%%%%%%%%%%%%%%%%%%%%%%%%%%%%
\section{Results}
%%%%%%%%%%%%%%%%%%%%%%%%%%%%%%%%%%%%%%%%%%%%%%%%%%%%%%%%%%%%%%%%%%%%%%%%%%%%%%%

For any theory with total quantum dimension $D$, one can compute the variational
energy per plaquette
%
%%%%%%%%%%%%%%%%
\begin{equation}
e(\alpha)=-\frac{J_\mathrm{p}f_\mathrm{p}(\alpha)
+3J_\mathrm{l}f_\mathrm{l}(\alpha)}{g(\alpha)},
\label{eq:gse_var}
\end{equation}
%%%%%%%%%%%%%%%%
%
where
%
%%%%%%%%%%%%%%%%
\begin{eqnarray}
g(\alpha)&=&D^2\left[D^2(1-\alpha)^2+4\alpha\right]^2,\\
f_\mathrm{p}(\alpha)&=&D^2(1+\alpha)^2\left[D^2(1-\alpha)^2+4\alpha\right],\\
f_\mathrm{l}(\alpha)&=&D^6(1-\alpha)^4+8D^4\alpha(1-\alpha)^3\nonumber\\
&&+24D^2\alpha^2(1-\alpha)^2+16\alpha^3(2-\alpha).
\end{eqnarray}
%%%%%%%%%%%%%%%%
%
Details of the calculations are given in  Appendix \ref{app_SN}. Setting $x=J_\mathrm{l}/J_\mathrm{p}$, the study of $e(\alpha)$ indicates that the system undergoes a  phase transition at $x_\mathrm{c}=\frac{D^2-1}{3D^2}$. Indeed, the minimum of $e$ is obtained for $\alpha_-=1$ if  $x\leqslant x_\mathrm{c}$ and for $\alpha_+\leqslant 1$ if  $x \geqslant x_\mathrm{c}$ (see Figs.~\ref{fig:e1}-\ref{fig:e2} for illustration). At the transition, one has \mbox{$\alpha_+(x_\mathrm{c})=\frac{D^2}{3D^2-4}$}. This transition is second order for $D^2=2$ only, and first order for \mbox{$D^2>2$}.

Interestingly, all these variational results only depend on~$D$. This is
reminiscent of the intrinsically local character of the ansatz that does not
take into account subtle effects due to nontrivial braiding statistics. Within
this mean-field approach (single-plaquette approximation), two theories with the
same total quantum dimension $D$ are thus treated on an equal footing.
Nevertheless, from high-order series expansions, we know that, for instance,
$\mathbb{Z}_4$ and Ising theories ($D^2=4$) have different ground-state energies
\cite{Schulz_unpub}.

Consequently, it is natural to wonder how these predictions compare with exact results. First, it is worth noting that, in the topological phase ($x< x_\mathrm{c}$), the energy is minimized for $\alpha=1$ which is the exact result for $x=0$. For $\alpha=1$, the ground-state energy reads 
%
%%%%%%%%%%%%%%%%
\begin{eqnarray}
\frac{e(\alpha_-)}{J_\mathrm{p}}&=&-1-\frac{3 x}{D^2},
\end{eqnarray}
%%%%%%%%%%%%%%%%
%
which matches the exact small-$x$ perturbative expansion up to order 1 but does
not give higher-order corrections. Secondly, in the opposite (large-$x$) limit,
the variational energy per plaquette can be expanded in powers of $1/x$ and
reads, at order 4,
%
%%%%%%%%%%%%%%%%
\begin{eqnarray}
\frac{e(\alpha_+)}{J_\mathrm{l}}&=&-3-\frac{1}{x}\frac{1}{D^2}-\frac{1}{x^2}\frac{D^2-1}{6  D^4}-\frac{1}{x^3}\frac{D^4-3D^2+2}{36  D^6} \nonumber \\
&&-\frac{1}{x^4}\frac{2D^6-11D^4+19D^2-10}{432 D^8}. 
\end{eqnarray}
%%%%%%%%%%%%%%%%
%
This expansion matches the exact large-$x$ series expansion up to order 3 but not beyond. Once again, this is due to the local character of the ansatz that does not capture quantum fluctuations beyond a single plaquette. 

%
%
%%%%%%%%%%%%%%%%%%%%%%%
\begin{table}[t]
	\begin{tabular}{|c|c|c|c|c|}
	\hline
	  & $\mathbb{Z}_2$&$\mathbb{Z}_3$ & Fibonacci & Ising\\
	\hline
	$D^2$ & 2 & 3 & $3.618$ & 4 \\
	\hline
	$x_\mathrm{c}$ (mean field) & 0.1667 & 0.2222 & 0.2412 & 0.25 \\
	\hline
	$x_\mathrm{c}$ (series)& 0.2097 \cite{He90} & 0.2466 \cite{Hamer92} & 0.261 \cite{Schulz13}& 0.267 \cite{Schulz14} \\
	\hline
	\end{tabular}
	\caption{Position of the transition point for several theories computed with the mean-field ansatz (\ref{eq:state}) and with series expansions.
	}
	\label{tab:xc}
\end{table}
%%%%%%%%%%%%%%%%%%%%%%%
%
%

Another important remark concerns the behavior of the so-called Wilson loop operators denoted $W^s_{\mathcal{C}_n}$ for a contour $\mathcal{C}_n$ enclosing $n$ plaquettes and a string of type $s$ (see Appendix \ref{app_SN}). In the deconfined (confined) phase, the expectation value of $W^s_{\mathcal{C}_n}$ is expected to scale as the perimeter (area) of $\mathcal{C}_n$ \cite{Wilson74}. Remarkably, the present mean-field approach displays this behavior since 
%
%%%%%%%%%%%%%%%%
\begin{equation}
\langle W^s_{\mathcal{C}_n}\rangle_\alpha= \varkappa_s d_s \left[ \frac{4\alpha}{D^2(1-\alpha)^2+4\alpha} \right]^n,
\label{eq:wilson}
\end{equation}
%%%%%%%%%%%%%%%%
%
where $\varkappa_s$ and $d_s$ are the Frobenius-Schur indicator and the quantum
dimension of the string $s$, respectively (see Appendix \ref{app_SN}).  In the topological
phase,  one has $\langle W^s_{\mathcal{C}_n}\rangle_{\alpha_-}=\varkappa_s d_s$
for any $\mathcal{C}_n$, which can be interpreted as a trivial perimeter law
with an infinite characteristic length. By contrast, in the polarized phase,
one has $\langle W^s_{\mathcal{C}_n}\rangle_{\alpha_+}=\varkappa_s d_s
\mathrm{e}^{-n / \mathcal A}$ where the characteristic area $\mathcal A$ is
readily obtained from Eq.~(\ref{eq:wilson}). 

Let us now compare the mean-field predictions with existing results. As
explained above, for the $\mathbb{Z}_N$ theory ($D^2=N$), the model is
equivalent to the $N$-state Potts model in a transverse field on the triangular
lattice. This model is known to display a second-order transition (Ising
universality class) for $N=2$, and a first-order transition for $N\geqslant 3$
(see Ref.~\cite{Wu82} for a review). Thus, the present mean-field treatment gives the correct order of the transition.
In Table \ref{tab:xc}, we give the position of the transition point
$x_\mathrm{c}$ obtained from series expansions and from the present mean-field
ansatz. Quantitatively, the difference between the results of both approaches
decreases as $D^2$ increases. For the Potts model, the mean-field theory is even
known to be exact for large $D^2=N$ \cite{Pearce84}. In this limit, one obtains
a first-order transition at $x_\mathrm{c}=1/3$.
Since $e(\alpha)$ only depends on $D$, this large-$D$ mean-field result is
expected to hold for all theories. 

However, for non-Abelian theories with finite $D$, the situation is more complex. In two recent studies \cite{Schulz13,Schulz14}, using series expansion and exact diagonalizations, it has been claimed that the phase transition for Fibonacci and Ising theories is second order but the present mean-field approach predicts a first-order transition ($D^2>2$). Although none of these methods are exact, we strongly believe that a (weakly) first-order scenario is correct. 
Apart from the mean-field result, this conclusion relies on two observations which have been overlooked. 
%
%
%%%%%%%%%%%%%%%%%%%%%%%
\begin{figure}[t]
\includegraphics[width=0.7\columnwidth]{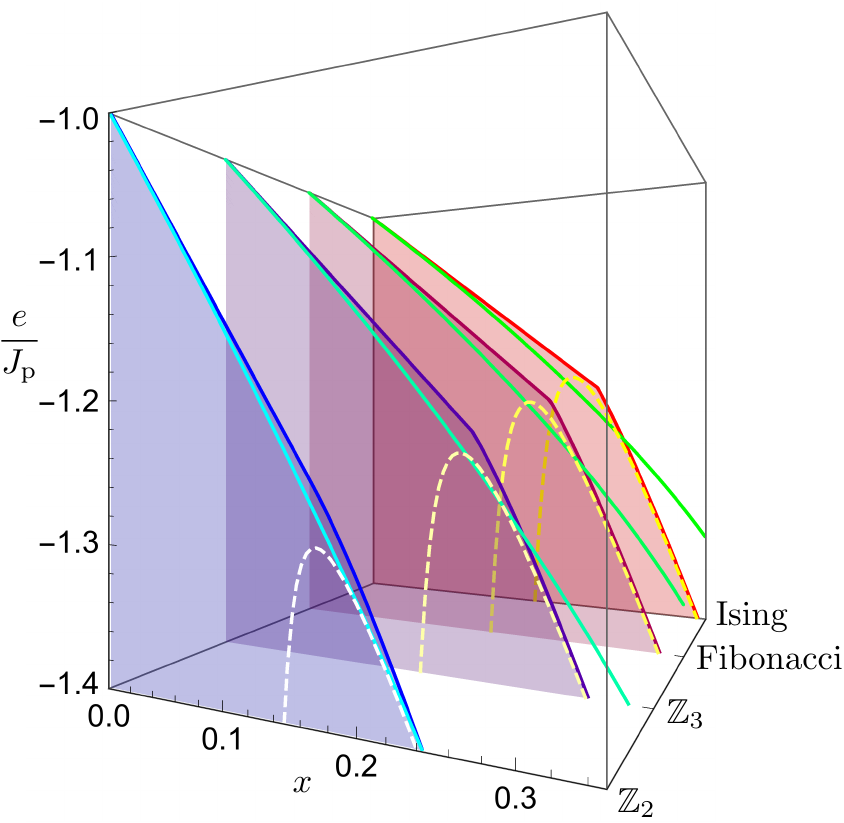}
\caption{(Color online) Comparison of variational results (upper boundaries of shaded planes) with low- (high-) field series expansions shown in full (dashed) lines for theories discussed in \mbox{Table \ref{tab:xc}}. Bare series at highest available orders \cite{He90,Hamer92,Schulz13,Schulz14} are displayed.
}
\label{fig:comparison}
\end{figure}
%%%%%%%%%%%%%%%%%%%%%%%
%
%

The first one relies on strong similarities of the ground-state energy series
expansions between $\mathbb{Z}_3$, Fibonacci, and Ising theories (see
Fig.~\ref{fig:comparison}). In particular,  a jump in the first derivative of
the ground-state energy per plaquette $\partial e / \partial x$ is observed at
the transition. This jump was considered as an artifact due to a finite-order
series in Refs.~\cite{Schulz13,Schulz14}. As can be seen in
Fig.~\ref{fig:comparison}, the magnitude of this jump is found to increase with
$D$, in agreement with the mean-field result which yields 
%
%%%%%%%%%%%%%%%%
\begin{equation}
\frac{\partial e}{\partial x}\Big|_{x=x_{\mathrm c}^-} -  \frac{\partial e}{\partial x}\Big|_{x=x_{\mathrm c}^+}= \frac{3(D^2-2)^2}{D^2(D^2-1)}.
\end{equation}
%%%%%%%%%%%%%%%%
%
%
In addition, for $D^2>2$, the (relative) height of the energy barrier at
$x=x_{\mathrm c}$ between the two minima $\alpha_\pm$ and the local maximum
$\alpha^*$ reads 
%
%%%%%%%%%%%%%%%%
\begin{equation}
\frac{e(\alpha_\pm)-e(\alpha^*)}{e(\alpha_\pm)}=\frac{D^4-4 D^2 \sqrt{D^2-1}+4D^2-4}{4(D^4+D^2-1)}.
\end{equation}
%%%%%%%%%%%%%%%%
%
%
In the limit $(D^2-2) \ll 1$, this relative energy vanishes as \mbox{$(D^2-2)^4$}, indicating a weakly first-order transition. 
Such a behavior qualitatively explains why the transition for Fibonacci and Ising theories has
been considered as second order in Refs.~\cite{Schulz13,Schulz14}.

The second argument that corroborates this scenario is based on the emergence of bound states in the low-energy spectrum inside the topological phase and will be discussed elsewhere. Let us simply mention that such bound states are necessary although not sufficient to induce a first-order transition and they are present for $D^2>2$.

The mean-field approximation can also be used to analyze the same model but in
the ladder geometry for which several exact results are known
\cite{Gils09_1,Gils09_3,Ardonne11,Schulz15}. In this one-dimensional case, the
variational energy is straightforwardly obtained from (\ref{eq:gse_var}) by merely 
replacing $J_\mathrm{l}$ by $J_\mathrm{l}/3$. As for the two-dimensional case,
the ansatz (\ref{eq:state}) predicts a second-order transition for $D^2=2$ and a
first-order transition for $D^2>2$ although, for the ladder, the transition is
known to be first order only if $D^2>4$ \cite{Schulz15}. Thus, the ansatz fails
at describing the nature of the transition for $D^2\leqslant 4$, as already
known for the Potts model \cite{Wu82}.
Interestingly, the position of the mean-field transition point
$x_\mathrm{c}^{\mathrm{ladder}}=\frac{D^2-1}{D^2}$ goes to 1 (self-dual point)
in the large-$D$ limit, which is the exact result for any $D$ \cite{Schulz15}.

%%%%%%%%%%%%%%%%%%%%%%%%%%%%%%%%%%%%%%%%%%%%%%%%%%%%%%%%%%%%%%%%%%%%%%%%%%%%%
\section{Conclusion}
%%%%%%%%%%%%%%%%%%%%%%%%%%%%%%%%%%%%%%%%%%%%%%%%%%%%%%%%%%%%%%%%%%%%%%%%%%%%%%%

To conclude, we would like to give some possible routes to go beyond the present approach. In Refs.~\cite{Gu09,Buerschaper09},
the ground state of the string-net model without string tension has been
written as a tensor-network state (TNS), involving a triple-line structure (that
reduces to a double-line structure for Abelian theories).
Following the steps detailed in these works, the state $\ket{\alpha}$ could be
written in the same way. The parameter $\alpha$ would only change the
values taken by the tensors. Since $\ket{\alpha}$ already captures
semiquantitatively the physics of the transition induced by string tension, it
seems reasonable to assume that performing a minimization over all parameters of
the tensors should give more precise results. The first-order nature of the
phase transition for $D^2>2$ should furthermore be favorable to the obtention of accurate results.

TNS have already been successfully used to study phase transitions in Abelian
models \cite{Gu08,Dusuel11,Schulz12,Liu15}. However, they have not yet been
applied to the more challenging non-Abelian models, although the principles for
doing so have been laid down \cite{Pfeifer10}. The technique exhibited in the
present paper can be considered as a first step, even though the
tensor-network structure has been bypassed. Let us emphasize that
single-parameter TNS have already been proposed for $\mathbb{Z}_2$ models
\cite{Gu08,Liu15}, but their single-line structure leads to qualitatively wrong
results (first-order transition). To solve this problem, Gu {\it et al.} introduced multiparameter double-line tensors \cite{Gu08}. It seems that the double-line structure (or triple-line structure for non-Abelian theories) is crucial since it encodes information about plaquettes that is necessary for an area law in the confined phase. Our single-parameter ansatz supports this conclusion.

Let us also stress that tensors must be chosen carefully, in order to allow for
topological states \cite{Chen10, Buerschaper13, Swingle10, Schuch10}. For the
toric code in a parallel magnetic field, we have shown (see Appendix \ref{app_TCF}
for a detailed calculation) that the topological entropy
\cite{Kitaev06_2,Levin06} vanishes for $\alpha<1$, i.e., in the
polarized phase, but is equal to $-\log_2 D$ (which is equal to $-1$ since
$D=2$) for $\alpha=1$, i.e., in the topological phase. We conjecture
that the same relations hold for the string-net model, for any theory. We leave
the calculation of the topological entropy, or of other measures \cite{Orus14_1,
Orus14_2} for future works.

The use of TNS would furthermore allow one to study other transitions. For instance, for the Fibonacci theory \cite{Schulz13},
the ground state for $J_\mathrm{p}=0$ and $J_\mathrm{l}<0$ is the state~$\ket{1}$ where all links carry a string $1$, namely, a Fibonacci anyon. The
transition from the string-net ground state to this state could thus be studied
with a variational state $\ket{\alpha}=\mN\prod_p(\id+\alpha Z_p)\ket{1}$.
However, analytical calculations are much harder in this case, so that numerical TNS methods would be extremely valuable.

Finally, let us mention that it would be interesting to describe excitations in
a variational setting and thus to study dynamical properties in the model, as was done
in Ref.~\cite{Levin07}. We hope the present work will trigger such studies.

%------------------------------------------------------------------------------
\acknowledgments
%------------------------------------------------------------------------------

We thank F. J. Burnell, R. Or\'us, K. P. Schmidt, M. D. Schulz, and S. H. Simon for fruitful discussions.

\appendix

%
%
%%%%%%%%%%%%%%%%
\section{String-net model with string tension}
\label{app_SN}
%%%%%%%%%%%%%%%%
%
%

%------------------------------------------------------------------------------
\subsection{Definitions}
%------------------------------------------------------------------------------
%
The Hamiltonian of the string-net model with a string tension is given by
%
%%%%%%%%%%%%%%%%
\begin{equation}
H= - J_\mathrm{p}\sum_p \Pp -J_\mathrm{l}\sum_l\Pe,
\end{equation}
%%%%%%%%%%%%%%%%
%
%
where the string-tension operator $\Pe$ is the projector onto the trivial state ${\ket 0}_l$ on the link $l$. 
The operator $\Pp$ enforcing trivial flux in plaquette $p$ is written as $\Pp=\frac{1}{D^2}\sum_{s}  \widetilde{d_s} B_p^s$.  Here, we introduce $\widetilde{d_s}=\varkappa_s d_s$, which is the product of the Frobenius-Schur indicator $\varkappa_s$ and of the quantum dimension $d_s$ of the string~$s$. The total quantum dimension is defined as $D=\sqrt{\sum_s d_s^2}$.  The operator $B_p^s$ inserts a string $s$ in the links of plaquette $p$ as defined in Appendix C of Ref.~\cite{Levin05}. 
Since $B_p^0$ acts as the identity on states satisfying branching rules (to which we restrict ourselves), we shall single it out and write $B_p^0=\id$. We introduce the operator $Z_p=2\Pp-\id$ that satisfies $Z_p^2=\id$ since $\Pp$ is a projector. With these notations and noting that $\widetilde{d}_0=1$, one obtains
%
%%%%%%%%%%%%%%%%
\begin{equation}
Z_p=-\frac{D^2-2}{D^2}\id+\frac{2}{D^2}\sum_{s\neq 0}\widetilde{d_s} B_p^s.
\label{eq:Zp}
\end{equation}
%%%%%%%%%%%%%%%%
%

%------------------------------------------------------------------------------
\subsection{Normalization of $\ket{\alpha}$}
%------------------------------------------------------------------------------
We consider the variational state
%
%%%%%%%%%%%%%%%%
\begin{equation}
\ket{\alpha}=\mN\prod_p(\id+\alpha Z_p)\ket{0},
\end{equation}
%%%%%%%%%%%%%%%%
%
where $0\leqslant\alpha\leqslant 1$ is a variational parameter. The fully polarized state ${\ket 0}$ is defined as ${\ket 0}= \otimes_l {\ket 0}_l$.
The first task is to compute the normalization constant $\mN$. 
To this end, let us note that
%
%%%%%%%%%%%%%%%%
\begin{equation}
(\id+\alpha Z_p)^2=(1+\alpha^2)(\id+\eta Z_p), \,\mbox{ with }\, \eta=\frac{2\alpha}{1+\alpha^2}.
\end{equation}
%%%%%%%%%%%%%%%%
%
Thus, denoting $\Np$ the number of plaquettes and since all $B_p^s$ commute
with one another, we find
%
%%%%%%%%%%%%%%%%
\begin{equation}
1=\braket{\alpha}=\mN^2(1+\alpha^2)^{\Np}
\bra{0}\prod_p\left(\id+\eta Z_p\right)\ket{0}.
\label{eq:N}
\end{equation}
%%%%%%%%%%%%%%%%
%
For simplicity and since we are interested in the thermodynamical limit,
let us assume open boundary conditions. Then, the only
contribution to $\bra{0}\prod_p\left(\id+\eta Z_p\right)\ket{0}$ comes from
the term proportional to $\id$ that arises when expanding
$\prod_p\left(\id+\eta Z_p\right)$. Indeed, the action of a $B_p^{s\neq 0}$ on $\ket{0}$,
for a boundary plaquette, introduces nontrivial strings in the boundary
links that cannot be compensated by any other operator $B_{p'\neq p}^{s'}$. Using Eq.~(\ref{eq:Zp}), it is then easy to get the normalization condition
%
%%%%%%%%%%%%%%%%
\begin{equation}
1=\braket{\alpha}=\mN^2(1+\alpha^2)^{\Np}\varepsilon^{\Np},
\label{eq:N2}
\end{equation}
%%%%%%%%%%%%%%%%
%
with
%
%%%%%%%%%%%%%%%%
\begin{equation}
\varepsilon=1-\eta\frac{D^2-2}{D^2}.
\end{equation}
%%%%%%%%%%%%%%%%
%

%------------------------------------------------------------------------------
\subsection{Computation of $\langle \Pp \rangle_\alpha$}
%------------------------------------------------------------------------------
Let us pick a particular plaquette $p$ and compute
$\langle \Pp \rangle_\alpha
=\bra{\alpha}\Pp\ket{\alpha}$.
Since all $B_p^s$ commute with one another, we get
%
%%%%%%%%%%%%%%%%
\begin{equation}
\langle \Pp \rangle_\alpha=\mN^2(1+\alpha^2)^{\Np}
\bra{0}\Pp\prod_{p'}\left(\id+\eta Z_{p'}\right)\ket{0}.
\end{equation}
%%%%%%%%%%%%%%%%
%
From the definition of $Z_p$, it is easy to derive the identity
$\Pp(\id+\eta Z_p)=(1+\eta)\Pp$. The prefactor of
$\id$ in this term is $\frac{1+\eta}{D^2}$. Proceeding along the same lines as
for the normalization of $\ket{\alpha}$, and using the expression of $\mN$
stemming from Eq.~(\ref{eq:N2}), we then find
%
%%%%%%%%%%%%%%%%
\begin{equation}
\langle \Pp \rangle_\alpha
=\frac{1+\eta}{D^2}\frac{1}{\varepsilon}.
\label{eq:expectPp}
\end{equation}
%%%%%%%%%%%%%%%%
%

%------------------------------------------------------------------------------
\subsection{Computation of $\langle \prod_p \Pp\rangle_\alpha$}
%------------------------------------------------------------------------------
Let $\mathcal{P}_n$ be a set of $n$ plaquettes. The same argument as above shows
that all plaquettes of $\mathcal{P}_n$ will have a contribution
$\frac{1+\eta}{D^2}\frac{1}{\varepsilon}$, while other plaquettes have a
contribution~1. As a consequence
%
%%%%%%%%%%%%%%%%
\begin{equation}
\Big\langle \prod_{p \in \mathcal{P}_n}  \Pp \Big\rangle_\alpha
=\left(\frac{1+\eta}{D^2}\frac{1}{\varepsilon}\right)^n
=\langle \Pp \rangle_\alpha^n.
\end{equation}
%%%%%%%%%%%%%%%%
%

%------------------------------------------------------------------------------
\subsection{Computation of $\langle \Pe \rangle_\alpha$}
%------------------------------------------------------------------------------
Let us finally turn to the computation of
\mbox{$\langle \Pe \rangle_\alpha
=\bra{\alpha}\Pe\ket{\alpha}$}, which is a
little bit more involved. We denote $p_1$ and $p_2$ the plaquettes sharing link
$l$. Then $\Pe$ commutes with all $Z_p$ operators, except those
acting at plaquettes $p_1$ and $p_2$. As a consequence
%
%%%%%%%%%%%%%%%%
\begin{widetext}
\begin{equation}
\langle \Pe \rangle_\alpha
=\mN^2(1+\alpha^2)^{\Np-2}
\bra{0}
(\id+\alpha Z_{p_1})(\id+\alpha Z_{p_2})
\Pe
(\id+\alpha Z_{p_1})(\id+\alpha Z_{p_2})
\prod_{p\neq p_1,p_2}(\id+\eta Z_p)
\ket{0}
\end{equation}
\end{widetext}
%%%%%%%%%%%%%%%%
%
As for the previous two computations, the only contribution to the matrix
element $\bra{0}\cdots\ket{0}$ comes from
the term proportional to $\id$ after expanding the operators. Consequently, we can already take
into account the contribution of $\prod_{p\neq p_1,p_2}(\id+\eta Z_p)$,
that is, $\varepsilon^{\Np-2}$ as well as the expression of $\mN$, stemming from
Eq.~(\ref{eq:N}), to write
%
%%%%%%%%%%%%%%%%
\begin{equation}
\langle \Pe \rangle_\alpha
=\left(\frac{1}{1+\alpha^2}\right)^2\frac{1}{\varepsilon^2}
\bra{\psi}\Pe\ket{\psi},
\label{eq:delta_l_1}
\end{equation}
%%%%%%%%%%%%%%%%
%
where
%
%%%%%%%%%%%%%%%%
\begin{equation}
\ket{\psi}=(\id+\alpha Z_{p_1})(\id+\alpha Z_{p_2})\ket{0}.
\end{equation}
%%%%%%%%%%%%%%%%
%
Denoting $\id+\alpha Z_p=\beta\id+\gamma C_p$, with
%
%%%%%%%%%%%%%%%%
\begin{equation}
\beta=1-\alpha\frac{D^2-2}{D^2}, \,\,\,
\gamma=\frac{2\alpha}{D^2}, \,\,\,\mbox{and}\,\,\,
C_p=\sum_{s\neq 0}\widetilde{d_s} B_p^s,
\end{equation}
%%%%%%%%%%%%%%%%
%
one gets
%
%%%%%%%%%%%%%%%%
\begin{equation}
\ket{\psi}=\left[\beta^2\id
+\beta\gamma\left(C_{p_2}+C_{p_1}\right)
+\gamma^2C_{p_1}C_{p_2}\right]\ket{0}.
\end{equation}
%%%%%%%%%%%%%%%%
%
Since $\Pe$ enforces a trivial ($s=0$) flux in link $l$, and since~$C_p$ operators introduce non-trivial fluxes, only the first and third terms of $\ket{\psi}$
contribute to the matrix element appearing in Eq.~(\ref{eq:delta_l_1}).
When acting with $C_{p_1}C_{p_2}$ on $\ket{0}$, the only way to obtain a trivial
flux in link $l$ is to take the same $s$ in $C_{p_1}$ and in $C_{p_2}$. Thus,
one gets all possible states with a loop $s$ surrounding plaquette~$p_1$ and a loop~$s$ surrounding
$p_2$. Since one requires the link $l$ to be in the trivial $s=0$ state, the weight
of these states is equal to~$\widetilde{d_s}$ as can be found by using Eq.~(2.23) in Ref.~\cite{Bonderson_thesis}.
As a consequence, we obtain
%
%%%%%%%%%%%%%%%%
\begin{equation}
\bra{\psi}\Pe\ket{\psi}
=\beta^4+\gamma^4\sum_{s\neq 0} \widetilde{d_s}^2
=\beta^4+\gamma^4(D^2-1),
\end{equation}
%%%%%%%%%%%%%%%%
%
so that
%
%%%%%%%%%%%%%%%%
\begin{equation}
\langle \Pe \rangle_\alpha
=\left(\frac{1}{1+\alpha^2}\right)^2\frac{1}{\varepsilon^2}
\left[\beta^4+\gamma^4(D^2-1)\right].
\label{eq:delta_l_2}
\end{equation}
%%%%%%%%%%%%%%%%
%

%------------------------------------------------------------------------------
\subsection{Computation of $e(\alpha)=\langle H\rangle_\alpha/\Np$}
%------------------------------------------------------------------------------
Finally, we can compute the variational energy per plaquette
%
%%%%%%%%%%%%%%%%
\begin{equation}
e(\alpha)=\frac{\bra{\alpha} H\ket{\alpha}}{N_p}=
-J_\mathrm{p}\langle \Pp \rangle_\alpha
-3J_\mathrm{l}\langle \Pe \rangle_\alpha,
\end{equation}
%%%%%%%%%%%%%%%%
%
where the factor of 3 comes from the fact that on a honeycomb lattice, the number of links is three times the number of plaquettes.  Replacing $\langle \Pp \rangle_\alpha$ and $\langle \Pe \rangle_\alpha$ by their expressions, and simplifying everything, one gets the energy per plaquette given in the main text.
Note that for the ladder geometry \cite{Gils09_1}, the link operator~$\Pe$ only acts on rungs. As there are as many rungs as plaquettes, the variational energy per plaquette for the ladder reads
$e^\mathrm{ladder}(\alpha)=-J_\mathrm{p}\langle \Pp \rangle_\alpha-J_\mathrm{l}\langle \Pe \rangle_\alpha$.\\

%------------------------------------------------------------------------------
\subsection{Computation of $\langle W^s_{\mathcal{C}_n}\rangle_\alpha$}
%------------------------------------------------------------------------------
For a contour $\mathcal{C}_n$ enclosing $n$ plaquettes, the Wilson loop operator $W^s_{\mathcal{C}_n}$ inserts a string $s$ along $\mathcal{C}_n$. In principle, one should consider two distinct operators, depending whether the string lies above or below the lattice. However, for our ansatz state, these operators have identical expectation values so that we denote both of them as~$W^s_{\mathcal{C}_n}$. This operator is given by $W^s_{\mathcal{C}_n}=\varkappa_s \mathcal{W}^s_{\mathcal{C}_n}$ where $\mathcal{W}^s_{\mathcal{C}_n}$ is the type-$s$ simple-string operator defined in Ref.~\cite{Levin05}, and is nothing but a multi-plaquette version of $B_p^s$. As $W^s_{\mathcal{C}_n}$ commutes with all $B_p^{s'}$ operators,
%
%%%%%%%%%%%%%%%%
\begin{equation}
\langle W^s_{\mathcal{C}_n} \rangle_\alpha=\varkappa_s\mN^2(1+\alpha^2)^{\Np}
\bra{0}\prod_p\left(\id+\eta Z_p\right)\mathcal{W}^s_{\mathcal{C}_n}\ket{0}.
\end{equation}
%%%%%%%%%%%%%%%%
%
Since $\mathcal{W}^s_{\mathcal{C}_n}\ket{0}$ is the state with a string $s$ along ${\mathcal{C}_n}$, the only non-zero contribution comes from $(N_p-n)$ operators~$\id$ for plaquettes outside ${\mathcal{C}_n}$, and from $n$ operators~$B_p^{\bar{s}}$ inside ${\mathcal{C}_n}$, annihilating the string $s$ (where $\bar{s}$ is the dual string of $s$). Each of the $n$ fusions of $s$ and $\bar{s}$ gives a factor $\varkappa_s/d_s$, and the resulting contractible $s$ loop gives a factor $d_s$. As a result
%
%%%%%%%%%%%%%%%%
\begin{equation}
\langle W^s_{\mathcal{C}_n} \rangle_\alpha=\varkappa_s\mN^2(1+\alpha^2)^{\Np}
\varepsilon^{\Np-n}\left(\frac{2\eta\widetilde{d_s}}{D^2}\right)^n\left(\frac{\varkappa_s}{d_s}\right)^n d_s.
\end{equation}
%%%%%%%%%%%%%%%%
%
Simplifying this expression finally yields
%
%%%%%%%%%%%%%%%%
\begin{equation}
\langle W^s_{\mathcal{C}_n} \rangle_\alpha=\varkappa_sd_s\left(\frac{2\eta}{D^2\varepsilon}\right)^n.
\end{equation}
%%%%%%%%%%%%%%%%
%

Let us mention that, as for a single plaquette, one can build the projector $W_{\mathcal{C}_n}=\frac{1}{D^2}\sum_s \widetilde{d_s}W^s_{\mathcal{C}_n}$ onto flux $0$ inside~$\mathcal{C}_n$. This operator
has the following expectation value:
%
%%%%%%%%%%%%%%%%
\begin{equation}
\langle W_{\mathcal{C}_n} \rangle_\alpha=\frac{1}{D^2}\left[1+(D^2-1)\left(\frac{2\eta}{D^2\varepsilon}\right)^n\right].
\end{equation}
%%%%%%%%%%%%%%%%
%
From this expression, it follows that $\langle W_{\mathcal{C}_n}
\rangle_{\alpha=1}=1$ as expected. Furthermore, when $n=1$, one can check that
$\langle W_{\mathcal{C}_{n=1}} \rangle_\alpha=\langle \Pp \rangle_\alpha$ given
in Eq.~(\ref{eq:expectPp}). This result allows one to rewrite the expectation value
of Wilson operators as:
%
%%%%%%%%%%%%%%%%
\begin{equation}
\langle W^s_{\mathcal{C}_n} \rangle_\alpha=\varkappa_s d_s
\left(\frac{D^2\langle \Pp \rangle_\alpha-1}{D^2-1}\right)^n.
\end{equation}
%%%%%%%%%%%%%%%%
%

%
%
%%%%%%%%%%%%%%%%%%%%%%%%%%%%%%%%%%%%%%%%%%%%%%%%%%%%%%%%%%%%%%%%%%%%%%%%%%%%%%%
\section{Toric code in a magnetic field}
\label{app_TCF}
%%%%%%%%%%%%%%%%%%%%%%%%%%%%%%%%%%%%%%%%%%%%%%%%%%%%%%%%%%%%%%%%%%%%%%%%%%%%%%%
%

%
%%%%%%%%%%%%%%%%%%%%%%
\subsection{Definitions}
%%%%%%%%%%%%%%%%%%%%%%
%

The Hamiltonian of the toric code in a magnetic field reads
%
%%%%%%%%%%%%%%%%
\begin{equation}
H=-J\sum_v A_v-J\sum_p B_p-\sum_l\boldsymbol{h}\cdot\boldsymbol{\sigma}_l,
\end{equation}
%%%%%%%%%%%%%%%%
%
where $\boldsymbol{h}=(h_x,h_y,h_z)$ is a uniform magnetic field and
$\boldsymbol{\sigma}_l=(\sigma^x_l,\sigma^y_l,\sigma^z_l)$ are Pauli operators
at link $l$ of a square lattice. Furthermore, $A_v=\prod_{l\in v}\sigma^x_l$ and
$B_p=\prod_{l\in p}\sigma^z_l$, where $v$ and $p$, respectively, denote vertices
and plaquettes of the lattice (see Ref.~\cite{Dusuel11}, and references therein
for a detailed discussion of this model). These operators all commute with one
another and $A_v^2=B_p^2=\id$. Note that, with these
definitions, $A_v$ and $B_p$ are not defined as projectors.  In the following,
we shall only consider a system in the thermodynamical limit with open boundary
conditions.

%
%%%%%%%%%%%%%%%%%%%%%%
\subsection{Ansatz and limiting cases}
%%%%%%%%%%%%%%%%%%%%%%
%
For a given direction of the magnetic field $\boldsymbol{h}$, following the mean-field prescription previously detailed for string nets, we introduce the following variational state:
%%
%%%%%%%%%%%%%%%%
\begin{equation}
\ket{\alpha,\beta}=\mN\prod_v(\id+\alpha A_v)\prod_p(\id+\beta B_p)
\ket{\boldsymbol{h}},
\label{eq:state_TC}
\end{equation}
%%%%%%%%%%%%%%%%
%
%
where $\ket{\boldsymbol{h}}$ denotes the state fully polarized in the field direction. When $\boldsymbol{h}=0$, the exact ground state is given by $\alpha=\beta=1$, whereas for $J=0$, it is obtained for $\alpha=\beta=0$.
Although the normalization constant $\mN$ is hard to compute for arbitrary $\ket{\boldsymbol{h}}$, it is possible to find exact expressions for some particular field directions. 

In the following, we focus on two simple directions: the parallel-field case where the field points in the $z$ (or equivalently $x$) direction \cite{Trebst07,Hamma08,Vidal09_1}, and the transverse-field case  where it points in the $y$ direction. In the former case, a second-order phase transition in the Ising universality class is known to occur for  \mbox{$h_z/J\simeq 0.328$} \cite{Trebst07,Hamma08,Vidal09_1}, whereas in the latter case, a first-order transition occurs for \mbox{$h_y/J=1$} \cite{Vidal09_2}. As we will see, the ansatz state $\ket{\alpha,\beta}$ qualitatively captures these two very different behaviors.

%%%%%%%%%%%%%%%%%%%%%%%%%%%%%%%%%%%%%%%%%%%%%%%%%%%%%%%%%%%%%%%%%%%%%%%%%%%%%%%
\subsection{Parallel field}
%%%%%%%%%%%%%%%%%%%%%%%%%%%%%%%%%%%%%%%%%%%%%%%%%%%%%%%%%%%%%%%%%%%%%%%%%%%%%%%
For a start, we consider a field \mbox{$\boldsymbol{h}=(0,0,h)$} pointing along the
$z$ axis. When $J$ vanishes, the ground state is
$\ket{\!\!\Uparrow}=\otimes_l\ket{\!\!\uparrow}_l$, namely, the polarized state
where all spins point in the $z$ direction. In the opposite limit where $h=0$,
the system is in the topological (toric code) phase. The ground state is then an eigenstate of all
$A_v$ and~$B_p$ operators, with eigenvalues $1$, that can be written
$\mathcal{N}\prod_v\left(\frac{\id+A_v}{2}\right)\ket{\!\!\Uparrow}$.  For $h\neq 0$, the Hamiltonian still commutes with all $B_p$ operators and the problem can then be mapped onto an Ising lattice gauge theory on the square lattice \cite{Trebst07}.
The ground state is an eigenstate of all $B_p$'s with eigenvalues $1$ which enforces $\beta=1$ in Eq.~(\ref{eq:state_TC}). Thus, we consider the following simple ansatz state:
%
%%%%%%%%%%%%%%%%
\begin{equation}
\ket{\alpha}=\ket{\alpha,\beta=1}=\mN\prod_v(\id+\alpha A_v)\ket{\!\!\Uparrow}.
\end{equation}
%%%%%%%%%%%%%%%%
%
The structure of this state is simple enough to allow for straightforward calculations of all quantities appearing in the Hamiltonian. 

%------------------------------------------------------------------------------
\subsubsection{Normalization}
%------------------------------------------------------------------------------
Since $A_v^2=1$, one has $(\id+\alpha A_v)^2=(1+\alpha^2)(\id+\eta A_v)$, where
%
%%%%%%%%%%%%%%%%
\begin{equation}
\eta=\frac{2\alpha}{1+\alpha^2}.
\end{equation}
%%%%%%%%%%%%%%%%
%
For a finite-size system with $N_\mathrm{v}$ vertices and open
boundary conditions, the normalization condition thus reads
%
%%%%%%%%%%%%%%%%
\begin{equation}
1=\braket{\alpha}=\mN^2(1+\alpha^2)^{N_\mathrm{v}}
\bra{\Uparrow\!\!}\prod_v(\id+\eta A_v)\ket{\!\!\Uparrow}.
\end{equation}
%%%%%%%%%%%%%%%%
%
The only contribution to
$\bra{\Uparrow\!\!}\prod_v(\id+\eta A_v)\ket{\!\!\Uparrow}$ arises from the term
proportional to $\id$ (i.e., that does not involve any $A_v$ operator), since an
$A_v$ operator for a boundary vertex flip boundary spins. These spin flips
cannot be compensated by the action of other $A_v$ operators. As a consequence,
the state $\ket{\alpha}$ is normalized if the following condition holds:
%
%%%%%%%%%%%%%%%%
\begin{equation}
1=\braket{\alpha}=\mN^2(1+\alpha^2)^{N_\mathrm{v}}.
\label{eq:mN_TC}
\end{equation}
%%%%%%%%%%%%%%%%
%

%------------------------------------------------------------------------------
\subsubsection{Computation of $\langle B_p\rangle_\alpha$ and
	$\langle A_v\rangle_\alpha$}
%------------------------------------------------------------------------------
Since all $B_p$ and all $A_v$ operators commute, it is easy to see that
%
%%%%%%%%%%%%%%%%
\begin{equation}
\langle B_p\rangle_\alpha=\bra{\alpha}B_p\ket{\alpha}=1.
\end{equation}
%%%%%%%%%%%%%%%%
%

The computation of $\langle A_v\rangle_\alpha=\bra{\alpha}A_v\ket{\alpha}$ is
also straightforward since
%
%%%%%%%%%%%%%%%%
\begin{eqnarray}
\langle A_v\rangle_\alpha&=&\mN^2(1+\alpha^2)^{N_\mathrm{v}}
\bra{\Uparrow\!\!}A_v\prod_{v'}(\id+\eta A_{v'})\ket{\!\!\Uparrow}\nonumber\\
&=&\bra{\Uparrow\!\!}(\eta\id+A_v)\prod_{v'\neq v}(\id+\eta A_{v'})
	\ket{\!\!\Uparrow}.
\end{eqnarray}
%%%%%%%%%%%%%%%%
Here, we used the normalization condition (\ref{eq:mN_TC}) and the fact
that $A_v^2=\id$. As for the calculation of the norm, the only nonzero
contribution comes from the term proportional to $\id$, so that one gets
%
%%%%%%%%%%%%%%%%
\begin{equation}
\langle A_v\rangle_\alpha=\eta.
\end{equation}
%%%%%%%%%%%%%%%%
%
This result is independent of $N_\mathrm{v}$ and is thus valid in the
thermodynamical limit. This will be the case for all quantities discussed below.

%------------------------------------------------------------------------------
\subsubsection{Computation of $\langle \prod_{v\in\mathcal{V}_n}A_v\rangle_\alpha$}
%------------------------------------------------------------------------------
Let $\mathcal{V}_n$ be a set of $n$ vertices. The same argument as above shows
that all vertices of $\mathcal{V}_n$ have a contribution $\eta$ while other
vertices have a contribution $1$, so that
%
%%%%%%%%%%%%%%%%
\begin{equation}
\Big\langle \prod_{v\in\mathcal{V}_n}A_v\Big\rangle_\alpha=\eta^n
	=\langle A_v\rangle_\alpha^n.
\label{eq:prodAs}
\end{equation}
%%%%%%%%%%%%%%%%
%
This factorization property illustrates the mean-field character of the variational state $\ket{\alpha}$.
%------------------------------------------------------------------------------
\subsubsection{Computation of $\langle\sigma_l^z\rangle_\alpha$}
%------------------------------------------------------------------------------
We now turn to the calculation of
$\langle\sigma_l^z\rangle_\alpha=\bra{\alpha}\sigma_l^z\ket{\alpha}$ at link
$l$:
%
%%%%%%%%%%%%%%%%
\begin{equation}
\langle\sigma_l^z\rangle_\alpha=\mN^2\bra{\Uparrow\!\!}
\prod_{v}(\id+\alpha A_v)\sigma_l^z\prod_{v}(\id+\alpha A_v)
\ket{\!\!\Uparrow}.
\end{equation}
%%%%%%%%%%%%%%%%
%
We denote $v_1$ and $v_2$ the two vertices that share link $l$. Then
$\sigma_l^z A_{v_j}=-A_{v_j}\sigma_l^z$ for $j=1,2$, while $\sigma_l^z$ commutes
with all other $A_v$ operators. Using the trivial identity
$(\id+\alpha A_{v_j})(\id-\alpha A_{v_j})=(1-\alpha^2)\id$, we obtain
%
%%%%%%%%%%%%%%%%
\begin{equation}
\langle\sigma_l^z\rangle_\alpha=\mN^2\bra{\Uparrow\!\!}
\prod_{v\neq v_1,v_2}(\id+\eta A_v)(1-\alpha^2)^2\ket{\!\!\Uparrow}.
\end{equation}
%%%%%%%%%%%%%%%%
%
We thus get
%
%%%%%%%%%%%%%%%%
\begin{equation}
\langle\sigma_l^z\rangle_\alpha=\left(\frac{1-\alpha^2}{1+\alpha^2}\right)^2.
\end{equation}
%%%%%%%%%%%%%%%%
%

%------------------------------------------------------------------------------
\subsubsection{Computation of $\langle\sigma_l^x\rangle_\alpha$ and
$\langle\sigma_l^y\rangle_\alpha$}
%------------------------------------------------------------------------------
For the sake of completeness, let us mention that
%
%%%%%%%%%%%%%%%%
\begin{equation}
\langle\sigma_l^x\rangle_\alpha=\langle\sigma_l^y\rangle_\alpha=0,
\end{equation}
%%%%%%%%%%%%%%%%
%
which follows from the fact that $\sigma_l^x$ and $\sigma_l^y$ both flip a
single spin and that this single spin flip cannot be compensated by any product
of $A_v$ operators.

%------------------------------------------------------------------------------
\subsubsection{Computation of the energy per link $e$}
%------------------------------------------------------------------------------
On the square lattice, in the thermodynamical limit, the number of plaquettes
$N_\mathrm{p}$ equals the number of vertices $N_\mathrm{v}$, and this number is
half the number of links $N_\mathrm{l}$ of the lattice. As a consequence, the
variational energy per link $e(\alpha)=\bra{\alpha}H\ket{\alpha}/N_\mathrm{l}$
can be written as follows if one gathers all previous results
%
%%%%%%%%%%%%%%%%
\begin{equation}
e(\alpha)=-\frac{J}{2}(\eta+1)-h\left(\frac{1-\alpha^2}{1+\alpha^2}\right)^2.
\end{equation}
%%%%%%%%%%%%%%%%
%
Denoting $\eta=\frac{2\alpha}{1+\alpha^2}=\cos\theta$ and
$\frac{1-\alpha^2}{1+\alpha^2}=\sin\theta$, one finally obtains
%
%%%%%%%%%%%%%%%%
\begin{equation}
e(\theta)=-\frac{J}{2}(\cos\theta+1)-h\sin^2\theta.
\end{equation}
%%%%%%%%%%%%%%%%
%

%------------------------------------------------------------------------------
\subsubsection{Analysis of the variational energy and phase diagram}
%------------------------------------------------------------------------------
The variational energy $e(\theta)$ exactly has the form one would obtain by (i) noting that the
Hamiltonian is dual to the transverse-field Ising model on the square lattice,
thanks to the duality transformation $A_v=\mu_v^z$ and
$\sigma_l^z=\mu_{v_1}^x\mu_{v_2}^x$ with $v_1$ and $v_2$ being the two adjoining
vertices to link $l$; (ii) performing a mean-field treatment from the
dual spin-$1/2$ variables $\mu$, namely, by writing the variational state as a product state
$\ket{\theta}=\otimes_v\ket{\theta}_v$ satisfying $\langle\mu_v^z\rangle_\theta=\cos\theta$,
\mbox{$\langle\mu_v^x\rangle_\theta=\sin\theta$}, and
$\langle\mu_{v_1}^x\mu_{v_2}^x\rangle_\theta=\sin^2\theta$.
Of course, a direct mean-field treatment based on the original variables
$\sigma$ cannot describe this transition since a product state is topologically
trivial.

The energy $e(\theta)$ can be studied easily. It has a single minimum at
$\theta=0$, i.e., $\alpha=1$, when \mbox{$x=h/J\leqslant x_\mathrm{c}=1/4$}. When
$x>x_\mathrm{c}$, a single minimum is found for
\mbox{$\cos\theta=x_\mathrm{c}/x$}, thus for $\alpha<1$.
This shows that there is a second-order quantum phase transition at
$x=x_\mathrm{c}$, between the low-field ($x<x_\mathrm{c}$) topological phase,
and the high-field ($x>x_\mathrm{c}$) polarized phase. 
The mean-field approach is thus able to capture the qualitative features of
the phase transition, since it is known that the transverse-field Ising model
has a second-order quantum phase transition at $x_\mathrm{c}\simeq 0.328$. However, the position of the critical point is about $24\%$ off  since we find $x_\mathrm{c}=1/4$. 

It is also interesting to note that, in the topological phase, 
$e(x<x_\mathrm{c})=-J$ which agrees with the order 1 perturbative expansion 
in the low-field limit $h/J\ll 1$. In the polarized phase, one has
%
%%%%%%%%%%%%%%%%
\begin{equation}
e(x>x_\mathrm{c})=-h-\frac{J}{2}-\frac{J^2}{16h},
\end{equation}
%%%%%%%%%%%%%%%%
%
which agrees with series expansion up to order 2 in the high-field limit $J/h \ll 1$. As explained for string nets, this is due to the fact that the mean-field ansatz only captures quantum fluctuations at a single-vertex level which is not sufficient to obtain the exact contributions at higher orders.  

%%%%%%%%%%%%%%%%%%%%%%%%%%%%%%%%%%%%%%%%%%%%%%%%%%%%%%%%%%%%%%%%%%%%%%%%%%%%%%%
\subsubsection{Topological entropy}
%%%%%%%%%%%%%%%%%%%%%%%%%%%%%%%%%%%%%%%%%%%%%%%%%%%%%%%%%%%%%%%%%%%%%%%%%%%%%%%

A reliable way to detect topological order in a given quantum state is to compute the topological entropy \cite{Kitaev06_2,Levin06}. In the following, we show that state $\ket{\alpha}$ has a nonvanishing topological entropy only if $\alpha=1$, which is in agreement with the nature of the phases expected on both sides of the critical point $x_\mathrm{c}$. 

\begin{figure}[t]
\includegraphics[width=4cm]{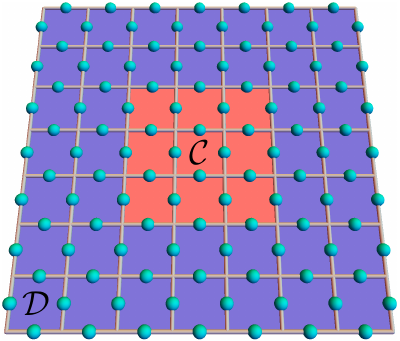}
\caption{(Color online) Partition of the lattice between two subsets $\mC$ and
$\mD$, where $\mC$ is simply connected.}
\label{fig:partition_CD}
\end{figure}

As shown in Ref.~\cite{Flammia09}, the topological entropy can be extracted from the computation of the R\'enyi
entanglement entropy.
We consider a system with open boundary conditions, and we split it into two
subsystems $\mC$ and $\mD$, where $\mC$ is simply
connected, as shown in Fig.~\ref{fig:partition_CD} for an example where
$\mC$ has a square shape.
We denote $c$, $d$, and $n$ the numbers of vertices fully included in $\mC$, 
fully included in $\mD$, and belonging to both $\mC$ and~$\mD$, respectively.
%
%
%%%%%%%%%%%%%%%%%%%%%%%
%%%%%%%%%%%%%%%%%%%%%%%
%
%
The aim is to compute the R\'enyi entanglement entropy between $\mC$ and
$\mD$ when the system is in state~$\ket{\alpha}$, namely,
$\mathcal{E}_2=-\log_2\left[\tr(\rho_\mC^2)\right]$, where
$\rho_\mC=\tr_\mD\ket{\alpha}\bra{\alpha}$. In the
thermodynamical limit, and for a domain $\mC$ that becomes bigger and
bigger, $\mathcal{E}_2=\beta n-\gamma+\cdots$ where $\cdots$ contains all terms
that vanish as $n\to\infty$. The first term is nonuniversal, contrary to the
second one $S_\mathrm{topo}=-\gamma$, which is the topological entropy. It can
be shown that $S_\mathrm{topo}=-\log_2 D$, where $D$ is the total quantum
dimension of the model under consideration. We shall prove below that
%
%%%%%%%%%%%%%%%%
\begin{equation}
S_\mathrm{topo}(\alpha=1)=-1 \,\, \mbox{ and } \,\,
S_\mathrm{topo}(0\leqslant\alpha<1)=0,
\end{equation}
%%%%%%%%%%%%%%%%
%
so that only the state $\ket{\alpha=1}$ has topological properties, with the
expected quantum dimension $D=2$, since there are four kinds of Abelian
particles in the toric code model~\cite{Kitaev03}.

In order to perform this calculation, we begin by rewriting $\ket{\alpha}$ as
follows:
%
%%%%%%%%%%%%%%%%
\begin{equation}
\ket{\alpha}=\frac{\mN}{\mN_\mC\mN_\mD}
\prod_{v\in\partial}(\id+\alpha A_v)\ket{\alpha}_\mC\otimes\ket{\alpha}_\mD,
\end{equation}
%%%%%%%%%%%%%%%%
%
where $\partial$ denotes the boundary of $\mC$ and $\mD$, namely, the vertices
belonging to both $\mC$ and $\mD$, and where
%
%%%%%%%%%%%%%%%%
\begin{eqnarray}
&&\ket{\alpha}_\mC=\mN_\mC\prod_{v\in\mC}(\id+\alpha A_v)\ket{\!\!\Uparrow}_\mC
\mbox{ with }
\ket{\!\!\Uparrow}_\mC=\otimes_{l\in\mC}\ket{\!\!\uparrow}_l,\qquad\\
&&\ket{\alpha}_\mD=\mN_\mD\prod_{v\in\mD}(\id+\alpha A_v)\ket{\!\!\Uparrow}_\mD
\mbox{ with }
\ket{\!\!\Uparrow}_\mD=\otimes_{l\in\mD}\ket{\!\!\uparrow}_l. \qquad
\end{eqnarray}
%%%%%%%%%%%%%%%%
%
These are normalized states, which impose the conditions
$\mN_\mC^2(1+\alpha^2)^c=1$ and $\mN_\mD^2(1+\alpha^2)^d=1$, that can be found
as before, since $\mC$ and $\mD$ have (at least) one boundary. Knowing that 
$\mN^2(1+\alpha^2)^{c+d+n}=1$, we can rewrite
%
%%%%%%%%%%%%%%%%
\begin{equation}
\ket{\alpha}=\mathcal{M}
\prod_{v\in\partial}(\id+\alpha A_v)\ket{\alpha}_\mC\otimes\ket{\alpha}_\mD,
\end{equation}
%%%%%%%%%%%%%%%%
%
with $\mathcal{M}^2(1+\alpha^2)^n=1$.

For each vertex $v\in\partial$, we write $A_v=A_v^\mC A_v^\mD$, where the
operator $A_v^\mC=\prod_{l\in v\cap\mC}\sigma_l^x$ flips the spins that are
belong to both $v$ and $\mC$, and where
$A_v^\mD=\prod_{l\in v\cap\mD}\sigma_l^x$ is defined similarly with domain
$\mD$. Then, state $\ket{\alpha}$ can be expanded as follows:
%
%%%%%%%%%%%%%%%%
\begin{eqnarray}
\ket{\alpha}&=&\mathcal{M}\Bigg(\ket{\alpha}_\mC\otimes\ket{\alpha}_\mD
+\alpha\sum_{v\in\partial}
	A_v^\mC\ket{\alpha}_\mC\otimes A_v^\mD\ket{\alpha}_\mD\nonumber\\
&&+\alpha^2\sum_{v_1\neq v_2\in\partial}
	A_{v_1}^\mC A_{v_2}^\mC\ket{\alpha}_\mC\otimes
	A_{v_1}^\mD A_{v_2}^\mD\ket{\alpha}_\mD+\cdots\nonumber\\
&&+\alpha^n
	A_{v_1}^\mC\cdots A_{v_n}^\mC\ket{\alpha}_\mC\otimes
	A_{v_1}^\mD\cdots A_{v_n}^\mD\ket{\alpha}_\mD\Bigg),
\end{eqnarray}
%%%%%%%%%%%%%%%%
%
where in the last term, all $v_1,\cdots,v_n$ are distinct and belong to
$\partial$.

For open boundary conditions, the states $\ket{\alpha}_\mD$,
$\left\{A_v^\mD\ket{\alpha}_\mD, v\in\partial\right\}$,
$\left\{A_{v_1}^\mD A_{v_2}^\mD\ket{\alpha}_\mD,
	v_1\neq v_2\in\partial\right\}$, $\cdots$,
$A_{v_1}^\mD\cdots A_{v_n}^\mD\ket{\alpha}_\mD$ are all different and
form an orthonormal set of states. It is thus easy to take the partial trace
needed to compute the reduced density matrix
$\rho_\mC=\tr_\mD\ket{\alpha}\bra{\alpha}$. One gets
%
%%%%%%%%%%%%%%%%
\begin{eqnarray}
\rho_\mC&=&\mathcal{M}^2\Bigg(\ket{\alpha}_\mC\,{}_\mC\bra{\alpha}
+\alpha^2\sum_{v\in\partial}A_v^\mC\ket{\alpha}_\mC\,
	{}_\mC\bra{\alpha}A_v^\mC+\cdots\nonumber\\	
&&+\alpha^{2n}A_{v_1}^\mC\cdots A_{v_n}^\mC\ket{\alpha}_\mC\,
{}_\mC\bra{\alpha}A_{v_1}^\mC\cdots A_{v_n}^\mC\Bigg),
\end{eqnarray}
%%%%%%%%%%%%%%%%
%
where we did not write as many terms as before to keep things as readable as
possible.

To compute $\tr(\rho_\mC^2)$, one needs to find the spectrum of $\rho_\mC$. For
this, one has to see that all states appearing in the expression of $\rho_\mC$,
namely, $\ket{\alpha}_\mC$,
$\left\{A_v^\mC\ket{\alpha}_\mC, v\in\partial\right\}$,
$\left\{A_{v_1}^\mC A_{v_2}^\mC\ket{\alpha}_\mC,
	v_1\neq v_2\in\partial\right\}$, $\cdots$,
$A_{v_1}^\mC\cdots A_{v_n}^\mC\ket{\alpha}_\mC$ are normed but not orthogonal to
each other, because one has the following identity
%
%%%%%%%%%%%%%%%%
\begin{equation}
A_{v_1}^\mC\cdots A_{v_n}^\mC=\prod_{v\in\mC}A_v.
\end{equation}
%%%%%%%%%%%%%%%%
%
Indeed, let us define two complementary states $\ket{\psi_1}_\mC$ and
$\ket{\psi_2}_\mC$, as states from the set written above that have the overlap
%
%%%%%%%%%%%%%%%%
\begin{eqnarray}
{}_\mC\langle\psi_1|\psi_2\rangle_\mC
&=& {}_\mC\bra{\alpha}A_{v_1}^\mC\cdots A_{v_n}^\mC\ket{\alpha}_\mC
= {}_\mC\bra{\alpha}\prod_{v\in\mC}A_v\ket{\alpha}_\mC \nonumber\\
&=& \eta^c,
\end{eqnarray}
%%%%%%%%%%%%%%%%
%
the last expression being obtained as Eq.~(\ref{eq:prodAs}).
Thus, two complementary states are not orthogonal. On the contrary, two states
that are not complementary are easily seen to be orthogonal.
As a consequence, the density matrix $\rho_\mC$ has a block diagonal structure.
Each block is a $2\times 2$ matrix involving two complementary states of the
form
%
%%%%%%%%%%%%%%%%
\begin{equation}
\rho_\mC^{(j)}=\mathcal{M}^2\left(
\alpha^{2j}\ket{\psi_1}_\mC\,{}_\mC\bra{\psi_1}
+\alpha^{2(n-j)}\ket{\psi_2}_\mC\,{}_\mC\bra{\psi_2}\right),
\end{equation}
%%%%%%%%%%%%%%%%
%
where $j$ is the number of $A_v^\mC$ operators appearing in
state $\ket{\psi_1}_\mC=A_{v_1}^\mC\cdots A_{v_j}^\mC\ket{\alpha}$, with
$0\leqslant j\leqslant n/2$. For the sake of simplicity, we shall consider that
$n$ is an even number, as is the case in Fig.~\ref{fig:partition_CD}.
When $j<n/2$, there are $\begin{pmatrix}n\\j\end{pmatrix}$ such
$\rho_\mC^{(j)}$ matrices. When $j=n/2$, there are
$\frac{1}{2}\begin{pmatrix}n\\n/2\end{pmatrix}$ matrices $\rho_\mC^{(n/2)}$.
 
The matrices $\rho_\mC^{(j)}$ can be rewritten in an orthonormal basis made of
states $\ket{\phi_1}_\mC$ and $\ket{\phi_2}_\mC$. For this, one first computes
the overlap matrix $\mathcal{O}$ with matrix elements
$\mathcal{O}_{k,l}={}_\mC\langle\psi_k|\psi_l\rangle_\mC$ with $k$ and $l$
taking values $1$ or $2$, namely,
%
%%%%%%%%%%%%%%%%
\begin{equation}
\mathcal{O}=\begin{pmatrix}
1 & \eta^c\\
\eta^c & 1
\end{pmatrix}.
\end{equation}
%%%%%%%%%%%%%%%%
%
This symmetric matrix can be diagonalized by performing a rotation,
$\mathcal{O}={}^\mathrm{t}\mathcal{P}\mathcal{D}\mathcal{P}$, with
%
%%%%%%%%%%%%%%%%
\begin{equation}
\mathcal{P}=\frac{1}{\sqrt{2}}\begin{pmatrix}
-1 & 1\\
1 & 1
\end{pmatrix}
\quad\mbox{and}\quad
\mathcal{D}=\begin{pmatrix}
1-\eta^c & 0\\
0 & 1+\eta^c
\end{pmatrix}.
\end{equation}
%%%%%%%%%%%%%%%%
%
With these definitions, and noting that the diagonal matrix $\mathcal{D}$ has
nonnegative diagonal elements, one can perform the change of basis
%
%%%%%%%%%%%%%%%%
\begin{equation}
\begin{pmatrix}
\ket{\psi_1}_\mC\\
\ket{\psi_2}_\mC
\end{pmatrix}
={}^\mathrm{t}\mathcal{P}\mathcal{D}^{1/2}\mathcal{P}
\begin{pmatrix}
\ket{\phi_1}_\mC\\
\ket{\phi_2}_\mC
\end{pmatrix}.
\end{equation}
%%%%%%%%%%%%%%%%
%
One can then express $\rho_\mC^{(j)}$ in the orthonormal basis of
states $\ket{\phi_1}_\mC$ and $\ket{\phi_2}_\mC$ (the expressions being quite
large, we shall not give them here).

A check of the validity of the obtained expression is to compute $\tr\rho_\mC$.
We find that
%
%%%%%%%%%%%%%%%%
\begin{eqnarray}
&&\tr\left(\rho_\mC^{(j)}\right)
=\frac{\alpha^{2j}+\alpha^{2(n-j)}}{(1+\alpha^2)^n},\\
\mbox{thus}\quad && \tr\rho_\mC=\frac{1}{2}
\sum_{j=0}^n\begin{pmatrix}n\\j\end{pmatrix}\tr\left(\rho_\mC^{(j)}\right)=1,
\end{eqnarray}
%%%%%%%%%%%%%%%%
%
as it should.
Note that we have extended the sum over $j=0,\cdots,n/2$ to
$j=0,\cdots,n$ and have corrected the induced double counting by the prefactor
$1/2$.

Similarly, one can compute $\tr\left({\rho_\mC^{(j)}}^2\right)$ and deduce
%
%%%%%%%%%%%%%%%%
\begin{eqnarray}
&&\tr\left(\rho_\mC^2\right)
=\frac{1}{2}\sum_{j=0}^n
\begin{pmatrix}n\\j\end{pmatrix}\tr\left({\rho_\mC^{(j)}}^2\right)\\
&&\qquad=\left[\frac{1+\alpha^4}{(1+\alpha^2)^2}\right]^n\left[1+
\left(\frac{2\alpha^2}{1+\alpha^4}\right)^n
\left(\frac{2\alpha}{1+\alpha^2}\right)^{2c}\right].\nonumber
\end{eqnarray}
%%%%%%%%%%%%%%%%
%
In the case $\alpha=1$, one finds $\tr\left(\rho_\mC^2\right)=2^{1-n}$, so that
$\mathcal{E}_2(\alpha=1)=n-1$ and $S_\mathrm{topo}=-1$, as
expected \cite{Kitaev06_2,Levin06}.
When $0\leqslant\alpha<1$, the second term in the above equation vanishes
exponentially fast when $n$ grows (for a generic domain $\mC$, $c$ grows like
$n^2$). One then gets the following behavior of $\mathcal{E}_2$:
%
%%%%%%%%%%%%%%%%
\begin{equation}
\mathcal{E}_2=n\log_2\left[\frac{(1+\alpha^2)^2}{1+\alpha^4}\right]+\cdots,
\end{equation}
%%%%%%%%%%%%%%%%
%
where $\cdots$ represents terms that vanish when taking the limit $n\to\infty$.
As a consequence, the topological entropy vanishes when $0\leqslant\alpha<1$.

%
%
%%%%%%%%%%%%%%%%%%
\subsubsection{Wilson loops}
%%%%%%%%%%%%%%%%%%
%
%

As already mentioned, the topological phase ($x<x_\mathrm{c}$) is the deconfined phase of the Ising lattice gauge model, in which Wilson loops are known to obey a perimeter law~\cite{Tagliacozzo11}. By contrast, in the polarized (deconfined) phase, these loops obey an area law. 
In the toric code model, Wilson loop operators $W_\mathcal{C}$ can be chosen as a product of $\sigma_l^x$ operators along a closed contour, which is nothing but the product of all operators $A_v$ surrounded
by this contour. Using Eq.~(\ref{eq:prodAs}), one thus gets $\langle W_\mathcal{C} \rangle_\alpha=\eta^n$ where $\eta=\cos\theta=\frac{2\alpha}{1+\alpha^2}$. In the topological phase, the energy is minimized for $\alpha=1$ so that $\langle W_\mathcal{C} \rangle_\alpha=1$ for any contour $\mathcal{C}$. This can be interpreted as a trivial
perimeter law with an infinite characteristic length. 
In the polarized phase ($\alpha<1$), one can write $\langle W_\mathcal{C} \rangle=\exp(-n\ln(1/\eta))$, which is an area law with a characteristic area $1/\ln(1/\eta)$. 
Our variational state thus correctly mimics the expected behavior of $W$ in the deconfined phase as well as in the confined phase.

%%%%%%%%%%%%%%%%%%%%%%%%%%%%%%%%%%%%%%%%%%%%%%%%%%%%%%%%%%%%%%%%%%%%%%%%%%%%%%%
\subsection{Transverse field}
%%%%%%%%%%%%%%%%%%%%%%%%%%%%%%%%%%%%%%%%%%%%%%%%%%%%%%%%%%%%%%%%%%%%%%%%%%%%%%%
We now consider a field $\boldsymbol{h}=(0,h,0)$ pointing along the $y$ axis.
In this case, the model is known to display a first-order quantum phase transition at the self-dual point $h=J$ \cite{Vidal09_2}.
The variational state reads
%
%%%%%%%%%%%%%%%%
\begin{equation}
\ket{\alpha,\beta}=\mN\prod_v(\id+\alpha A_v)\prod_p(\id+\beta B_p)
\ket{\!\!\Rightarrow},
\end{equation}
%%%%%%%%%%%%%%%%
%
where $\ket{\!\!\Rightarrow}=\otimes_l\ket{\!\!\rightarrow}_l$, is the polarized
state where all spins point in the $y$ direction. For the sake of completeness,
we introduced two variational parameters $\alpha$ and $\beta$ but, for symmetry reasons, we expect them to be equal. 

Since all calculations follow closely that of the preceding section, we shall directly give the results without further justification. 
The expectation values of charge and flux operators read
%
%%%%%%%%%%%%%%%%
\begin{equation}
\langle A_v\rangle_{\alpha,\beta}=\frac{2\alpha}{1+\alpha^2} \mbox{ and }
\langle B_p\rangle_{\alpha,\beta}=\frac{2\beta}{1+\beta^2}.
\end{equation}
%%%%%%%%%%%%%%%%
%

In addition, one has $\langle\sigma_l^x\rangle_{\alpha,\beta}=0$,  $\langle\sigma_l^z\rangle_{\alpha,\beta}=0$, and
%
%%%%%%%%%%%%%%%%
\begin{equation}
\langle\sigma_l^y\rangle_{\alpha,\beta}
=\left(\frac{1-\alpha^2}{1+\alpha^2}\right)^2
\left(\frac{1-\beta^2}{1+\beta^2}\right)^2.
\end{equation}
%%%%%%%%%%%%%%%%
%

Setting $\cos\theta=\frac{2\alpha}{1+\alpha^2}$, $\sin\theta=\frac{1-\alpha^2}{1+\alpha^2}$ and
$\cos\phi=\frac{2\beta}{1+\beta^2}$, $\sin\phi=\frac{1-\beta^2}{1+\beta^2}$, and keeping in mind that, in the
thermodynamical limit, $N_\mathrm{v}=N_\mathrm{p}=N_\mathrm{l}/2$, one gets the following energy per link:
%
%%%%%%%%%%%%%%%%
\begin{equation}
e(\theta,\phi)=-\frac{J}{2}\cos\theta-\frac{J}{2}\cos\phi
-h\sin^2\theta\sin^2\phi.
\end{equation}
%%%%%%%%%%%%%%%%
%

As expected, this variational energy is found to be minimum for
$\phi=\theta$, i.e., for $\alpha=\beta$. Again, the expression of $e(\theta,\theta)$ could have been
obtained by using a duality transformation and treating the dual model in a mean-field way. The study of $e(\theta,\theta)$ shows that $\theta=0$ is always a minimum, but
is the absolute minimum only for $x=h/J<x_\mathrm{c}=27/32$. For
$x\geqslant x^*=\frac{3\sqrt{3}}{8}$, a second local minimum appears and it
becomes the absolute minimum for $x>x_\mathrm{c}$. For $x=x_\mathrm{c}$, two
absolute minima coexist, at $\theta=0$ and at $\theta=\arccos(1/3)$. 
We thus find a first-order quantum phase transition at $x=x_\mathrm{c}$. 
Consequently, our variational analysis is thus about $16\%$ off the exact result $x_\mathrm{c}=1$, and misses the self-duality of the model \cite{Vidal09_2}.  

Finally, in the topological phase, $e(x<x_\mathrm{c})=-J$ agrees with the low-field series expansion up to order 1 in $h/J$. In the polarized phase, the series expansion of the variational energy at order 4 in $J/h$ reads %
%%%%%%%%%%%%%%%%
\begin{equation}
e(x<x_\mathrm{c})=-h-\frac{J^2}{8h}-\frac{J^4}{256h^3},
\end{equation}
%%%%%%%%%%%%%%%%
%
which matches the high-field series expansion up to order~2 in $J/h$ \cite{Vidal09_2} (odd order contributions vanish). 

%\bibliography{./biblio_LW}

%merlin.mbs apsrev4-1.bst 2010-07-25 4.21a (PWD, AO, DPC) hacked
%Control: key (0)
%Control: author (0) dotless jnrlst
%Control: editor formatted (1) identically to author
%Control: production of article title (0) allowed
%Control: page (1) range
%Control: year (0) verbatim
%Control: production of eprint (0) enabled
%

\end{document}